\documentstyle[11pt]{article}
\renewcommand{\baselinestretch}{1.5}
\newcommand{\add}{\addtocounter{eqncnt}{1}}
\newcommand{\med}[1]{g_{#1}}
\newcommand{\meu}[1]{g^{#1}}
\newcommand{\be}{\begin{equation}}
\newcommand{\ee}{\end{equation}\add}
\newcommand{\bea}{\begin{eqnarray}}
\newcommand{\eea}{\end{eqnarray}}
\newcommand{\ol}{\overline}
\renewcommand{\theequation}{\thesection.\theeqncnt}
\newcommand{\und}{\underline}
\newcommand{\Dsp}{\displaystyle}
\newcommand{\ens}{\enskip}
\begin{document}
\begin{center}{\Large \bf The Similarity Hypothesis in General Relativity} \\[2mm]
\vskip .5in{\sc B. J. Carr}\\{\em Astronomy Unit, Queen Mary, University of London}\\
{\em Mile End Road, London E1 4NS, England}\\[2mm]\em and \\[2mm] {\sc A.A. Coley}\\
{\it Department of Mathematics and Statistics}\\{\it Dalhousie University,Halifax, Nova ScotiaCanada  B3H 3J5}
\end{center}

\begin{abstract}

Self-similar models are important in general relativity and
other fundamental theories. In this paper we shall discuss the
``similarity hypothesis'', which asserts  that under a variety of
physical circumstances solutions of these theories will naturally
evolve to a self-similar form. We will find there is good evidence for this in the context of
both spatially homogenous and inhomogeneous cosmological models, although in some cases the self-similar model is only an intermediate attractor. There are also a wide variety of situations, including critical pheneomena, in which spherically
symmetric models tend towards self-similarity. However, this does not happen in all cases and it is important to understand the prerequisites for the conjecture.

\end{abstract}
\vskip .5in\section{Introduction}

In Newtonian hydrodynamics self-similar solutions occur when the physical quantities depend on
functions of $x / l (t)$, where $x$ and $t$ are independent space and time variables and $l$ is a
time-dependent scale.  This means that the spatial distribution of the characteristics of motion
remains similar to itself at all times during the motion and that all dimensional constants entering
the initial and boundary conditions vanish or become infinite (Barenblatt and Zeldovich 1972; BZ).
When the form of the self-similar asymptotics can be obtained from dimensional considerations, the
solutions are referred to as self-similar solutions of the {\it first} kind (BZ).  Otherwise, the solutions
are referred to as self-similar solutions of the more general {\it second} kind and in this case they
contain dimensional constants.  In general relativity the concept of self-similarity is less
straightforward, since there are various ways of generalizing the Newtonian concept and a covariant
characterization is required.  The exist ence of self-similar solutions of the first kind is
related to conservation laws and can be characterized by the existence of a homothetic vector.
However, in general the self-similar variables are not determined from dimensional analysis alone and
one then has self-similar solutions of the second kind.  As in the Newtonian context, a
characteristic of these solutions is that they contain dimensional constants which are not determined
from the conservation laws but can be found by matching the self-similar solutions with the
non-self-similar solutions whose asymptotes they represent (BZ). A more detailed discussion of the different types of self-similarity can be found in the review of Carr \& Coley (1999).

Self-similar models are of great interest in physics because they are often found to play an important role in in describing the asymptotic properties of more general models.  Historically, this was first noticed in the context of  fluid dynamics, where it was found that -- for a wide range of problems -- self-similar asymptoticscan be obtained from dimensional considerations. Indeed, it was in this context that the concept of self-similarity was first introduced (BZ).
Subsequent developments have shown that self-similar solutions may play a similar role in general relativity (Carr and Coley 1999).  For example, in the cosmological context  it seems that there are a wide variety of circumstances in which spherically symmetric solutions naturally  evolve to a self-similar form and this led Carr to  formulate what he termed the ``similarity
hypothesis'' (Carr 1993). This asserts that, under certain physical
circumstances, which have never been precisely specified, spherically symmetric solutions will naturally evolve to a self-similar
form.  Later it was found that self-similar solutions also play a crucial asymptotic role for spatially homogeneous cosmological models (Wainwright \& Ellis 1997; WE) and indeed in a variety of non-cosmological situations. Therefore the context of the similarity conjecture is now very broad.

In this paper we shall discuss the applicability of the similarity hypothesis in general relativity
and review the various circumstances in which it is known to hold.  We will focus mainly on
self-similarity of the first kind (i.e.,  homothetic similarity) but we will also comment on the more
general self-similarity of the second kind and on ``kinematic" self-similarity (Carter and
Henriksen 1989, Coley 1997).  In Section 2 we consider the evidence that spatially homogenous
cosmological models tend to self-similar form at early or late times.  In Section 3, we extend the
discussion to spatially inhomogeneous (but not spherically symmetric) models.  In Section 4 we discuss
solutions which approach self-similarity only at intermediate times (including inflation).  In Section
5 we review spherically symmetric self-similar solutions and in Section 6 we consider various
applications of these (inaked singularities and critical phenomena); we also include a discussion of their stability.  In Section 7
we mention the status of the similarity hypothesis for theories other than general relativity and in Section 8
we draw some general conclusions.

\section{Spatially Homogeneous Model} Spatially homogeneous models
have attracted considerable attention since the governing ordinary differential equations (ODEs)
reduce to a relatively simple dynamical system (Wainwright and Ellis 1997; WE).  Utilizing the
orthonormal frame method and introducing expansion-normalized (and hence dimensionless)
commutation-function variables and a new ``dimensionless'' time variable, the equations for fluid
models with equation of state $p = \alpha \rho$ reduce to a finite-dimensional system of coupled
autonomous ODEs.  For $-1<\alpha \le 1$, all of the singular points of this
reduced dynamical system have been shown to correspond to self-similar cosmological models (Hsu and
Wainwright 1986), which is why self-similar models play an important role in describing the asymptotic
dynamics of the Bianchi models at both early and late times.  The case $\alpha =1$, corresponding to a
stiff fluid, requires special consideration but is also fairly well understood.  This is related to
the case in which one has a massless scalar field, since such a field is equivalent to a stiff fluid
providing its gradient is timelike.

\subsection{Early Times} Let us first consider the asymptotic properties of spatially homogeneous
Bianchi models with $0 \leq \alpha < 1$ at early times.  The dynamics of these models have been
studied using dynamical systems methods (WE) and more rigorous techniques (Rendall 1997).  A large
class of such models -- including all those in class B (Bianchi types IV, V, VI$_h$, VII$_h$) -- are
asymptotically self-similar at the initial singularity and well approximated by exact perfect fluid or
vacuum power-law models.  The asymptotes include self-similar Kasner vacuum models and self-similar
locally rotationally symmetric Bianchi type II perfect fluid models (Collins and Stewart
1971, Collins 1971).

However,  orthogonal models of Bianchi type IX and VIII exhibit oscillatory behaviour as one goes back to the initial singularity, with the matter density becoming dynamically negligible. Such models have chaotic-like characteristics and are generically asymptotically Mixmaster rather than self-similar.
%The Mixmaster dynamics is described by a past
%attractor, which is the union of the Kasner circle of equilibrium points
%and the vacuum Bianchi II orbits (cf. Hobill et al., 1994).
The orbits of the associated dynamical system are negatively asymptotic to a lower two--dimensional attractor, which consists of the Kasner ring joined byTaub separatrices, and the orbits spend most of the time near the Kasner vacuum equilibrium points (cf. Hobill et al., 1994). Therefore, although these modelsdo not have a well-defined asymptotic state, they are approximated by aninfinite sequence of successiveself-similar models (Kasner vacuum solutions) as one goes back towardsthe initial singularity.
Perfect fluid models with $0 \leq \alpha < 1$ have subsequently
been analysed more rigorously by Ringstr\"om (2000, 2001). He has proved
the existence of the past attractor, using the Hubble-normalized evolution
equations, and confirmed the generic nature of the dynamics.

An oscillatory approach to
the initial singularity has also been exhibited in
magnetic Bianchi $VII_{0}$ cosmologies (Horwood and  Wainwright, 2004),
in non-tilted exceptional type VI$_{-1/9}$ Bianchi cosmologies (Hewitt et al., 2003),
and in tilted spatially homogeneous Bianchi cosmologies
(Hewitt et al., 2001; Coley and Hervik, 2005) for which the tilt is
dynamically significant. The initial singularity  (past attractor) has also been
studied numerically (Uggla et al. 2003).

In scalar field models, the scalar field is effectively
massless at early times and dynamically
equivalent to a stiff fluid providing the gradient of the scalar field is timelike (Coley 2003).
It is known that the evolution near the initial singularity in four-dimensional spacetimes with a
massless scalar field or stiff fluid matter is non--oscillatory and approximated by a
self-similar exact solution (Andersson and Rendall 2000).

\subsection{Late Times} Self-similar solutions can also describe the behaviour of Bianchi models at
late times (i.e.,  as $t \rightarrow \infty$).  All such models expand indefinitely except for
the Bianchi IX models, which obey the ``closed universe recollapse" conjecture (Lin and Wald
1989).  The vacuum models of Bianchi type IV, V, VI$_h$ and VII$_h$ are future-asymptotic to plane-wave states, with the type V models tending to the Milne form of flat spacetime (Hewitt
and Wainwright 1993).  Non-vacuum models are typically, and perhaps generically (Hewitt and
Wainwright 1993), asymptotic to either plane-wave vacuum solutions or non-vacuum Collins type
VI$_h$ solutions (Collins 1971, WE).

Non-tilted exceptional Bianchi cosmologies of
type VI$_{-1/9}$ are also asymptotically self-similar at late times; the future
attractor for vacuum-dominated
models is the so-called Robinson-Trautman model (Hewitt et al., 2003).
The asymptotic properties of stiff
perfect fluid models ($\alpha=1$) are also known (Wainwright 1998) and the Bianchi VIII models
are the only ever-expanding stiff models that do not have a self-similar future asymptote.  As
we have noted, this
is also relevant in the study of scalar field models (Coley 2000).

The asymptotic
dynamics of tilted $p=\alpha \rho$ spatially homogeneous cosmologies of Bianchi type II (Hewitt et al., 2001), type VI$_0$ (Barrow and Hervik, 2003, Hervik, 2004, Coley and Hervik, 2004)
and more general Bianchi models of solvable type (e.g., VI$_h$  and VII$_h$)
have also been discussed (Coley and Hervik, 2005; Hervik et al., 2005). It has been shown that for non-inflationary solutions of this type there are always future-stable vacuum plane-wave
solutions; in fact, these are the only future-attracting
equilibrium points in the Bianchi type VII$_h$ invariant set (Coley and Hervik, 2005).
A tiny region of parameter space
(the ``loophole") in the Bianchi type IV model was shown to contain a closed orbit which acts as an attractor. A detailed numerical analysis
of the type VII$_h$ models then shows that there is an open set of parameter space in which
solution curves approach a compact surface that is topologically a torus
(Hervik et al., 2005). In all of these models the future asymptotic state is self-similar.

The late time
behaviour of Bianchi type VII$_0$ models with a non--tilted perfect fluid source has also been
studied (Wainwright et al.  1999; Nilsson et al.  2000).  Since the Bianchi VII$_0$ state space
is unbounded, it can be shown that these
models are generally {\em not}
asymptotically self-similar at late times and the same applies for Bianchi type VIII models (Horwood and Wainwright, 2004).
This is because it is found that asymptotically there is  `Weyl curvature dominance', characterized by the
divergence of the Hubble-normalized Weyl curvature at late times,
if $0<\alpha <1 $ for Bianchi VII$_0$ models (Wainwright et al., 1999)
and if $0\leq \alpha < 1$ for Bianchi VIII models (Horwood and Wainwright 2004).
In these Bianchi models generic orbits escape to infinity,
such that the Weyl tensor diverges as $t\rightarrow +\infty$ and
is consequently dynamically dominant at late times, and there
is no future attractor and self-similarity breaking occurs (Horwood et al., 2003).
The future asymptote behaviour in Bianchi type VIII vacuum models was confirmed by a more rigorous
analysis (Ringstrom 2003).

In an investigation of the asymptotic dynamics of the Einstein-Maxwell field equations
for a class of Bianchi cosmologies with perfect fluid and
a pure homogeneous source-free magnetic field,
it was shown that typical magnetic Bianchi VII$_{0}$
cosmologies also exhibit Weyl curvature dominance in the late-time regime
(Horwood and  Wainwright, 2004).

The analysis of the qualitative properties of spatially homogeneous models with a non-tilted
perfect fluid source with $p=\alpha \rho$ has been extended in a
number of ways.
Tilting perfect fluid models of various Bianchi types have been comprehensively
studied (Hewitt and Wainwright 2000,
Hewitt et al., 2001, Coley and Hervik, 2005, Hervik et al., 2005), as have two-perfect-fluid models (Coley and Wainwright 1992) and imperfect fluid Bianchi models (Coley and van den Hoogen 1994, 1995).
Spatially homogeneous cosmological models with a magnetic field (and a non-tilted perfect fluid)
have also been studied (Leblanc 1998, Weaver 2000, Clarkson et al.  2001,
Horwood and Wainwright 2004).
In all cases, self-similar solutions have been found to play an important role in the asymptotic
behaviour of the models.

The asymptotic properties of spatially
homogeneous models with a scalar field have also been studied; in particular, in scalar field models
with an exponential potential, the future asymptotic
behaviour is again self-similar (Billyard et al.  1999, Coley 2003).

\section{Spatially Inhomogeneous (Non-Spherical) Models}
Self-similar solutions have also been found to play an
important role in describing the asymptotic properties of more general spatially inhomogeneous
models.  This includes silent-universe models (Bruni et al.  1995) and $G_2$ models (Hewitt et al.
1988, 1993), the latter admitting two commuting orthogonally transitive spacelike Killing vectors.
The spherically symmetric inhomogeneous models
are discussed separately in Section 4.  In such
studies, the Einstein field equations (regarded as evolutionary equations) are partial
differential equations and the resulting state space is thus an infinite-dimensional function
space.  A useful formalism for inhomogeneous models (including the spherically symmetric ones) has been provided by van Elst and Uggla (1997).

\subsection{Early Times} The most detailed proposal for the structure of initial
space--time singularities is that of Belinskii, Khalatnikov and Lifshitz (BKL).  This has the
following features.  (1) The matter can only be dynamically significant close to the singularity
if it is stiff (Lifshitz and Khalatnikov 1963, Belinskii et al.  1970).  (2) Each spatial point
evolves towards the singularity as if it were part of a spatially homogeneous cosmology, i.e.,
spatial points decouple near the singularity, the Einstein field equations effectively reducing to ODEs and the local dynamical behaviour being asymptotically Bianchi-like near
the singularity (Belinskii et al.  1971, 1982).  (3) The singularities in generic
four-dimensional spacetimes are spacelike and oscillatory for non-stiff fluids (Belinskii et al.
1971, 1982) but spacelike and non--oscillatory for stiff fluids and massless scalar fields
(Belinskii and Khalatnikov 1972).

The analysis in the spatially homogeneous case (discussed in Section 2.1) already supports the
BKL conjecture.  Studies of spatially inhomogeneous solutions near the initial singularity
extend this result to a more general context.  In particular, the spatially inhomogeneous $G_2$
models have been investigated.  Again normalized
variables are introduced and the evolution equations for the perfect fluid
models can be written as a first-order system of quasi-linear partial differential
equations.  The {\it equilibrium points} correspond to exact self-similar (and in general
inhomogeneous) $G_2$ models (WE).  In particular, the special class of self-similar diagonal
$G_2$ cosmologies, in which the governing equations reduce to a finite-dimensional invariant
subset, have been studied by Hewitt and Wainwright (1993).  They conjecture that self-similar
states may describe the asymptotic dynamical behaviour of spatially inhomogeneous $G_2$ models.  The qualitative analysis of these
diagonal models constitutes a tentative first step in the fundamental analysis of more general $G_2$ models.

Other special classes of Abelian $G_{2}$ spatially inhomogeneous models have been analysed.  In
the so-called ``velocity-dominated'' spacetimes, the evolution at different spatial points is
found to approach different Kasner solutions (Berger and Moncrief 1993 and 1998,
 Berger and
Garfinkle 1998, Kichenassamy and Rendall 1998).
This provides further support for the similarity hypothesis for spatially inhomogeneous
cosmological models at early times.
A numerical investigation of a class of $G_{2}$
cosmological spacetimes that are inhomogeneous generalizations of magnetic field Bianchi type
VI$_0$ models indicates that at most points the evolution toward the initial singularity is
asymptotic to a spatially homogeneous spacetime with Mixmaster behavior (Weaver et al.  1998).

There are special classes of models that do not evolve
according to the BKL conjecture but are still asymptoticaly self-similar.  The most important
perhaps are those containing an {\em isotropic initial singularity\/}, whose evolution near the
singularity is approximated by the flat FRW model (Goode
and Wainwright 1982, Goode et al. 1992).  These models are of interest for a variety
of physical reasons and are, of course, trivially self-similar at the singularity itself
(Penrose 1976).

\subsection{Late times}

Although attention has mainly focussed on the late-time behaviour
of spatially homogeneous cosmological models and their linear perturbations, there
are two obvious exceptions to this. First,
the cosmic no-hair theorems for  spatially homogeneous models (discussed in more detail in Section 4) have been generalized to spatially inhomogeneous models. For example,
recently there have been numerical investigations of the late-time behaviour
of generic inhomogeneous $G_0$ models (Lim et al. 2003), and the asymptotics of solutions of  ever-
expanding cosmological solutions with positive
cosmological constant have been studied rigorously (Rendall, 2003).
The cosmic no-hair theorems provide further evidence
that cosmological models are future asymptotic to exact self-similar models.

The second exception relates to the stability of plane-wave solutions.
We have seen in Section 2.1 that all orthogonal class B perfect fluid models,
with equation of state parameter $\alpha \le 1$, expand indefinitely
into the future and an open set of models
are future-asymptotic
to a self-similar vacuum plane wave  solution (WE).
It is of interest to know whether such models are stable to inhomogeneous perturbations.
As noted above, a detailed analysis of the tilted Bianchi models shows that
the plane-wave geometry is
asymptotically stable for types IV, VII$_h$, and an open set of type VI$_{h}$
models. Thus there exists an open set of spatially homogeneous
universes asymptoting towards the vacuum plane-waves at late times
(Coley and Hervik, 2005; Hervik et al., 2005).
It is also known from a linearised
perturbation analysis that the vacuum plane-wave spacetimes are stable
within a class of inhomogeneous  $G_2$ models -- see, for example, Barrow and Tsagas (2005) --
and in these models there does exist a late time self-similar asymptote. However,
it was recently shown (Hervik and Coley, 2005) that the plane-wave spacetimes of
Bianchi type VII$_h$ are unstable with respect to general inhomogeneous perturbations;
the critical unstable modes were found to correspond to
inhomogeneous perturbations orthogonal to the propagation of the
gravitational wave.
The asymptotic behaviour in particular inhomogeneous scalar field models has also been investigated;
self-similarity occurs in special classes of $G_2$ models (Coley, 2003).
% and spherically symmetric models (Coley and Goliath 2000, Coley and Taylor 2001, Coley and He 2002).

\section{Inflation and Intermediate Self-Similarity}

Many years ago Misner (1968) suggested that dissipative effects in the early Universe might drive
the Universe towards spatial homogeneity and isotropy, even if it starts out chaotic.
If there exists a physical mechanism which drives solutions towards this form, then this would also
support the similarity hypothesis.  In the context in which it was originally proposed, this
scenario turns out to be implausible since, with arbitrarily high initial chaos, more entropy would
be generated than is observed in the cosmic background radiation (Barrow and Tipler 1986).
% More appropriate to cite Stewart paper?
However, the scenario is much more plausible in the context of the popular inflationary scenario,
which might be viewed as a contemporary version of the chaotic paradigm.

Of course, inflationary models not only asympotote toward aflat self-similar FRWmodel -- they are also accelerating under the influence of an effective cosmological constant. In the case of spatially homogeneous models, it has been shown that allmodels with a cosmological constant are future-asymptotic to either the inflationary flat de Sitter model, which possesses self-similarity of the more general ``second" kind (Coley 1997), or
a flat self-similar power-law inflationary FRW model (Coley 2003). This is known as the cosmic no-hair theorem (Wald 1983) and provides further evidence that cosmological models are future-asymptoticto exact self-similar models.

A cosmological constant corresponds to an equation of state $\alpha =-1$ but the cosmic no-hair theorem has been extended to inflationary models with $-1 \leq \alpha < -1/3$ by
Coley and Wainwright (1992), who showed that all such spatially homogenous perfect fluid models are
future asymptotic to an exact flat self-similar inflationary FRW model.  Such asymptotic behaviour
is also common in dynamical scalar field models (Olive 1990).  In particular, the asymptotic
properties of spatially homogeneous models with a scalar field and an exponential potential have
been studied (Billyard et al.  1999, Coley 2000).  The singular points of the reduced dynamical
system correspond to exact self-similar solutions.  The cosmic no-hair theorem still applies and
the asymptotic states are again self-similar power-law inflationary FRW models.

After inflation has
driven the model towards spatially homogeneity and isotropy, physical mechanisms bring
inflation to an end and thereafter the model evolves away from this highly symmetric state.
However, observations of type IA supernovae at high redshift (Riess et al.  1998, Garnavich et al.
1998, Perlmutter et al.  1999, Riess et al.  2005) suggest that the Universe is currently accelerating.
This could be due to a cosmological constant or an exotic perfect fluid  with negative
pressure (dark energy) or a dynamical component whose energy density and spatial distribution evolve with time.
The last possibility motivates the so-called ``quintessence" models, which contain an evolving scalar field
(Caldwell et al., 1998, Ratra and Peebles, 1988), including those with an exponential potential
(Wetterich 1988, Copeland et al.  1998, Billyard et al.  1999).  Therefore self-similar behaviour
may still be physically important at late times

We have primarily discussed the applicability of the similarity hypothesis at early and late times.
However, self-similar solutions may also describe the ``intermediate-asymptotic'' behaviour of
solutions in a region in which they no longer depend on the details of the initial and/or boundary
conditions but in which the system is still far from equilibrium.  For example, in chaotic
cosmologies the Universe is driven towards a spatially homogeneous and isotropic state only for a
finite period.  Eventually it will evolve away from the highly symmetric state.  Similarly,
although inflation necessarily drives the Universe towards flatness, it is never exactly flat
unless it started out flat.  Therefore, the deviations from flatness will
eventually become significant again (Ellis 1988), so
%add Ellis reference
the self-similar flat models are only important at intermediate stages of evolution.

There are various other
examples of intermediate self-similarity.  In particular, exact self-similar power-law models can
approximate general Bianchi models at intermediate stages of their evolution.  Of special interest
are those models which can be approximated by an isotropic solution at an intermediate stage (e.g.,
models whose orbits spend a period of time near to the universal flat Friedmann equilibrium point)
and which are therefore important in relating Bianchi models to the real universe (WE, Wainwright
et al.  1998).

\section{Spherically Symmetric Models}

We have seen that the self-similarity assumption reduces the complexity of Einstein's equations. Even greater simplification is achieved in the case of spherical symmetry (Cahill \& Taub 1971), since the governing equations reduce to comparatively simple ODEs. In this case, the solutions can be put into a form in which every dimensionless variable is a function of some dimensionless combination of the time coordinate $t$ and the comoving radial coordinate $r$. In the simplest situation, a similarity solution is invariant under the transformation $r \rightarrow ar, \; t \rightarrow at$ for any constant $a$, so the similarity variable is $z=r/t$. Geometrically this corresponds to the existence of a homothetic vector and is termed similarity of the ``first'' kind. These solutions have been thoroughly investigated and are discussed in detailed in Section 5.2.

There are also similarity solutions of the ``second'' kind, which involve an intrinsic scale.
A particularly important application of this is {\it kinematic} self-similarity (Carter \& Henriksen 1989,  1991, Coley 1997), in which the similarity variable is of the form $z=r/t^\beta$ for some constant $\beta$.  For perfect fluid spacetimes of this kind, it has been
found that all spherically symmetric models asymptote at late times towards exact solutions that admit a
homothetic vector (Benoit and Coley 1998).  This provides evidence for the similarity hypothesis
in a broader context. Recently Maeda and colleagues have provided a fairly complete classification of kinematic self-similar solutions for $p=\alpha \rho$ fluids (Maeda et al. 2002a, 2002b, 2003).

These solutions should be distinguished from ones which possess {\it discrete} self-similarity,
in the sense that dimensionless variables repeat themselves on some spacetime scale
$\Delta$; i.e., they are invariant under the transformation $r \rightarrow e^{-n\Delta}r, \; t \rightarrow e^{-n\Delta}t$ for any integer $n$. One recovers {\it continuous} self-similarity in the limit $\Delta \rightarrow 0$. Although discrete self-similarity is much harder to deal with mathematically, Choptuik (1997) has emphasized its crucial role in the context of critical phenomena and this is discussed further in Section 6.2.

The first studies of spherically symmetric self-similar solutions focussed on the dust
case ($\alpha =0$) since these solutions can often be expressed analytically. References
to such solutions can be found in Kramer et al.  (1980) and Krasinski (1997).  In the
cosmological context,  such solutions are just a special subclass of the
more general spherically symmetric Lemaitre-Tolman-Bondi solutions (Tolman 1934, Bondi 1947, Bonnor
1956). Since the density on an initial spacelike surface may be chosen arbitrarily (Bonnor 1974), the general Lemaitre-Tolman-Bondi solution will generally {\it not} tend to self-similarity. Indeed,  the evolution of dust solutions is prescribed entirely by an ``energy function" $E(r)$ and
they will only tend to self-similarity if this function is constant (Carr 2000). On the other hand, there certainly exist dust models
which approach spatially homogeneous and isotropic FRW models at late times. For example, in the parabolic case, every Lemaitre-Tolman-Bondi solution asymptotes towards the Einstein de Sitter
model at late times, and a subclass of hyperbolic models also approaches the FRW model.
Although such models are asymptotically self-similar, they are generally not stable.
Note that Henriksen (1989) has argued that all spherically symmetric dust models in general
relativity are self-similar in the most general Lie sense, so that perfect fluid models that asymptote towards
a dust model always satisfy the similarity hypothesis in this sense.
%Surely this is not the relevant sense?

\subsection{Evidence of Similarity Hypothesis in Spherical Context}

What makes spherically symmetric self-similar solutions of more than mathematical interest is the fact that they are often relevant to the real world.  This is because, in a variety of astrophysical and
cosmological situations (both Newtonian and relativistic), solutions may naturally evolve to
self-similar form even if they start out more complicated.  This is illustrated by the following
examples.

* An explosion in a homogeneous background produces fluctuations which may be very
complicated initially but which tend to be described more and more closely by a Newtonian
spherically symmetric similarity solution as time evolves (Taylor 1950, Sedov 1967).  This applies even if
the explosion occurs in an expanding cosmological background (Schwartz et al.  1975, Ikeuchi et
al.  1983).

* The evolution of cosmic ``voids'' is also described by a similarity solution at
late times (Hoffman et al.  1983, Hausman et al.  1983, Bertschinger 1985).  Indeed Bertschinger
(1985) applied this idea to explain giant cosmic voids but found that one needs more than the
linear fluctuations which arise in the standard hierarchical clustering scenario.  Such voids
are usually discussed in a Newtonian context (since they are smaller than the particle horizon)
but they can also be studied in a relativistic context (Sato 1984, Tomita 1995, 1997).

* For all self-gravitating
systems, if the source of the motion can be considered as a point source, then a self-similar disturbance should result, with
self-similarity being a good approximation far from the source.
Explosions or voids are just particular examples of this, as are thermal waves (Zeldovich and Kompaneets 1950, Barenblatt 1952, Zeldovich and Raizer 1963).

* Numerical simulations show that
overdense regions in the hierarchically clustering scenario also tend to generate self-similar
cosmic structures (Quinn et al.  1986, Frenk et al.  1988, Efstathiou et al.  1988).  This is partly understood theoretically (Gunn and Gott 1972, Gunn 1977, Fillmore and
Goldreich 1984, Bertschinger and Watts 1988) but it is unclear to what extent it arises due to
non-linear effects rather than because the primordial fluctuations were scale-free.

* In the Newtonian context, a gravitationally bound cloud collapsing from an initially uniform static
configuration may evolve to a self-similar form (Larson 1969, Penston 1969, Hunter 1977, Hunter 1986, Lynden-Bell \& Lemos 1988).
This is not well understood theoretically, although it might be
associated with
such processes as virialization,
shell-crossing
and violent relaxation (Lynden-Bell-1967).

* In the relativistic context, numerical studies of quasi-static spherically symmetric
gravitational collapse also give strong evidence that the solutions tend
towards self-similar form.  In particular, Harada \& Maeda (2001) have studied the spherical collapse of a perfect fluid with equation of state $p=\alpha \rho$. For $0<\alpha<0.036$, there is a general relativistic counterpart of the Larson-Penston self-similar Newtonian solution.  Near the centre, generic collapse converges to this solution as it approaches the singularity.
As discussed in Section 6.1, such solutions are of particular interest because a naked singularity forms for $0<\alpha<0.0105$.
%this is the most serious known counter-example of cosmic censorship. It also provides strong evidence for the self-similarity hypothesis in general relativistic gravitational collapse.

* As discussed in Section 6.2, self-similar solutions are also relevant to the occurrence
of critical phenomena in gravitational collapse (Choptuik 1993, Evans and Coleman 1994,
Koike et al. 1995, Maison 1996, Carr et al.  1999, Gundlach 2003). Although the critical
solution only arises with very fine-tuned initial conditions, it is the
most stable solution  in the sense that it is only unstable to a single mode (Maison 1996).

* All static spherically symmetric perfect fluid solutions with $p=\alpha \rho$
are asymptotic at large distances to self-similar models (Collins 1985, Goliath et al.
1998b).  These are presumably associated with the asymptotically quasi-static model
discussed in Section 5.3.

These examples suggest that self-similar solutions may play the same sort of role in describing the asymptotic properties of spherically symmetric models as they do in the context of the spatially homogeneous solutions.  Indeed it was in this context that term  ``similarity hypothesis'' was first introduced, the proposal being that -- under certain circumstances (e.g., non-zero pressure, non-linearity, shell-crossing etc.) -- spherically symmetric solutions may naturally evolve to self-similar form (Carr 1993).

\subsection{Classification of spherically symmetric self-similar models}
The possibility that
self-similar models may be singled out in this way from more general spherically symmetric
solutions means that it is essential to understand the full family of self-similar solutions.  If
the source of the gravitational field is a single perfect fluid, Cahill \& Taub (1971) have shown that
the only equation of state compatible with the similarity assumption has the form $p = \alpha
\rho$.  In this case, it is well known that, for
a given value of $\alpha$, spherically symmetric similarity solutions are described by two
parameters.  Such solutions have now been classified completely, using both the
``comoving" approach (in which the coordinates are adapted to the fluid 4-velocity)
and the ``homothetic" approach (in which the coordinates are adapted to the homothetic vector).
These approaches have been taken by Carr \& Coley (2000) and Goliath et al. (1998), respectively, but a full understanding of the solutions
requires that one combines them (Carr et al. 2001). It is found that there are four classes of solutions and we now discuss these. However, it should be stressed this classification does not cover imperfect fluids or multiple fluids or solutions which contain shocks. The classification described below is therefore ``complete" only in a rather restricted sense.

* The first class consists of
the 1-parameter family of solutions asymptotic to the exact $k =0$ Friedmann model at large $z$
and these are obviously relevant in the cosmological context.  Some of them contain black holes
which grow at the same rate as the particle horizon (Carr \& Hawking 1974, Bicknell \& Henriksen
1978a).  Others represent density perturbations in a Friedmann background which always maintain
the same form relative to the particle horizon (Carr \& Yahil 1990).  The latter all contain a
sonic point and the requirement that they be regular there severely restricts their form
(Bogoyavlenski 1985).  While there is a continuum of regular underdense solutions, regular
overdense solutions only occur in narrow bands.

* The second class of models are asymptotic to a self-similar member of
the Kantowski-Sachs (KS) family.  For each non-zero $\alpha$, there is a unique self-similar KS
solution and there also exists a $1$-parameter family of solutions asymptotic to this at both
large and small values of $z$ (Carr \& Koutras 1992).  Solutions with $-1/3 < \alpha <1$ are
probably unphysical because they are tachyonic and the mass is negative.  Solutions with $-1<
\alpha < -1/3$ avoid these problems and are therefore more interesting.  Although this equation
of state violates the strong energy condition, it could could well arise in the early Universe.
Indeed, such models may be related to the growth of bubbles formed at a phase transition in an
inflationary model (Wesson 1986).  Generalized negative-pressure KS solutions, in which the
similarity is of the second kind, have been studied by Ponce de Leon (1988) and Wesson (1989);
in these the equation of state deviates from $p=\alpha \mu$ and the fluid can be interpreted as
a mixture of false vacuum and dust.

* The third class of models are associated with self-similar
static models.  There is just one static self-similar solution for each positive value of
$\alpha$ (Misner \& Zapolsky 1964) and there is also a 1-parameter family of solutions
asymptotic to this.  However, there is a 2-parameter family of solutions which are
asymptotically ``quasi-static'' in the sense that they have an isothermal density profile at
large values of $z$.  Some asymptotically quasi-static solutions have been
studied by Foglizzo \& Henriksen (1993).  These are of particular interest because they
are are associated with the formation of naked singularities for $\alpha < 0.0.036$ and the
occurrence of critical phenomena for $\alpha < 0.28$.

* The fourth class of solutions, which only exist for $\alpha
>1/5$, are asymptotically Minkowski.  They were originally found numerically by Goliath et al.
(1998) and this led Carr and Coley to analyse them analytically.  There are actually two such families
and they are considered in detail by Carr et al.  (2000).  Members of the first family are
described by one parameter and are asymptotically Minkowski as $|z|\rightarrow \infty$; members
of the second family are described by two parameters and are asymptotically Minkowski as $z$
tends to some finite value (though this corresponds to an infinite physical distance unless
$\alpha=1$).  As with the asymptotically Friedmann and asymptotically quasi-static solutions,
these may be either supersonic everywhere (in which case they contain a black hole or naked
singularity) or attached to $z=0$ via a sonic point (in which case they are asymptotically
Friedmann or exactly static at small $|z|$).  The transonic ones are associated with critical
phenomena for $\alpha > 0.28$ (Carr et al.  1999).

Many of the solutions discussed above possess a sonic point and a crucial feature of such solutions is that the equations do not determine their behaviour uniquely there. This is because there can be a number of different solutions passingthrough any sonic point. However, only a subset of solutions are ``regular''  in the sense that  they can be extended beyond the sonic point. Regular solutions can have just two values of the pressure gradient at each sonic point.  If both values are real, at least one of them will be positive.  If both values are positive (corresponding to a``nodal'' point), the smaller one is associated with a $1$-parameter family of solutions, while the larger oneis associated with an isolated solution.  If one of the values of isnegative (corresponding to a ``saddle'' point), both values are associatedwith isolated solutions. This behaviour has been analysed in detail byBogoyavlenski (1977), Bicknell and Henriksen (1978), Carr and Yahil (1990)and Ori and Piran (1990).On each side of the sonic point, the pressure gradient may have either of the two values.  If one chooses different values, there will be a discontinuity in the pressure gradient (corresponding to a sound-wave).  If one chooses the same value,there may still be a discontinuity in the higher derivatives of the pressure.  Onlythe isolated solution and a single member of the one-parameter family ofsolutions  are analytic. This contrasts with the case of a shock, where the pressureis itself discontinuous  (Cahill and Taub 1971).  It can be shown that there are two ranges of nodal sonic points, one including the static sonic point and the other the Friedmann sonic point.

\section{Applications of Spherically Symmetric Similarity Hypothesis }

\subsection{Gravitational
Collapse and Naked Singularities} Self-similarity is very relevant to the cosmic censorship
hypothesis because many of the known counterexamples involve exact homothetic solutions (Eardley
et al.  1986, Zannias 1991).  Indeed, it has been shown that a large subclass of self-similar
solutions have a central singularity from which null geodesics emerge to infinity (Henriksen and
Patel 1991) and it has been argued that one might generally expect a naked singularity to have a
horizon structure similar to that of the global homothetic solution (Lake 1992).  The occurrence
of naked singularities in spherically symmetric, perfect fluid, self-similar collapse has been
studied by Ori and Piran (1987, 1990), Waugh and Lake (1988, 1989), Lake and Zannias (1990),
Henriksen and Patel (1991), and Foglizzo and Henriksen (1993).  Their occurrence has also been studied in the context of a massless scalar field model by Christodoulou (1994) and Brady (1995) and in the context of the SU(2) sigma model by Bizon and Wasserman (2002).

Most of the early work focussed on dust solutions of the Lemaitre-Tolman-Bondi class, including
the analytical studies of Eardley and Smarr (1979) and Christodolou (1984).  Ori and
Piran (1990; OP) extended this work by studying spherically symmetric homothetic models with pressure.
For reasonable equations of state, it might be expected that pressure gradients would prevent the
formation of shell-crossing singularities (the situation is less clear for shell-focussing
singularities).  However, OP proved the existence of a ``significant'' class of perfect fluid
self-similar solutions with a globally-naked central singularity.  They explicitly studied the
causal nature of these solutions by analysing the equations of motion for the radial null
geodesics, thereby demonstrating that these geodesics emerge to infinity.  OP noted that
these perfect fluid (non-dust) solutions might constitute the strongest known counter-example to
cosmic censorship.
Foglizzo and Henriksen (1993) extended OP's analysis of the
gravitational collapse of homothetic perfect fluid gas spheres with $p = \alpha \rho$ for all
$\alpha$ between 0 and 1, partially utilizing the powerful dynamical systems approach of
Bogoyavlenski (1985).  They showed that the set of globally {\it analytic} naked solutions is
discrete but finite (and even empty for large values of $\alpha$) and they confirmed that the
number of oscillations in the flow is a good index, with the approach to the ``static'' solution
being recovered as this index grows.  Carr \& Gundlach (2003) have analysed the global structure of all the self-similar solutions found by CC and identified which ones have naked singularities.

The crucial question in the context of the similarity hypothesis is whether these solutions are
stable to perturbations.  As discussed in Section 6.3, recent work has indicated the stability of the naked singularity
solutions in both the massless scalar field case (Christodoulou 1999) and the perfect fluid case
(Harada 2001, Harada and Maeda 2004).  Indeed, the
solutions corresponding to both censored and naked singularities are topologically stable (Nolan
2001).  All of the self-similar solutions with naked singularities exhibit Cauchy horizons and a detailed
analysis of the stability of these horizons shows that they are stable with respect to scalar
radiation providing the fluid satisfies the dominant energy condition (Nolan and Waters 2002).

\subsection{Critical phenomena} One of the most exciting developments in general relativity in
the last decade has been the discovery of critical phenomena.  This first
arose in studying the gravitational collapse of a spherically symmetric massless (minimally
coupled) scalar field (Choptuik 1993).  If one considers a family of imploding scalar wave
packets whose strength is characterized by a continuous parameter $l$, one finds that the final
outcome is either gravitational collapse for $l > l^*$ or dispersal -- leaving behind a regular
spacetime -- for $l < l^*$.  For $(l-l^*)/l^*$ positive and small, the final black hole mass obeys
a scaling law $M_{BH} = C(l-l^*)^{\gamma}$ where C is a family-dependent parameter and
$\gamma=0.37$ is family-independent.  Initial data with $l = l^*$ evolve towards a critical
solution which exhibits ``echoing''.  This is a discrete self-similarity in which all
dimensionless variables $\Psi$ repeat themselves on ever-decreasing spacetime scales:
$\Psi(t,r) = \Psi(e^{-n \Delta} t, e^{-n \Delta} r)$ where n is a positive integer and
$\Delta=3.44$.  Solutions with near-critical initial data first evolve towards the critical solution, showing
some echoing on small space scales, but then rapidly deviate from it to either form a black
hole ($l > l^*$) or disperse ($l < l^*$).

Obtaining analytical results with a
discrete self-similarity is difficult, so attempts have been made to elucidate critical
phenomena by studying spherically symmetric solutions which possess continuous self-similarity.
Spherically symmetric homothetic spacetimes containing radiation were first investigated by Evans
\& Coleman (1994).  They studied models containing ingoing Gaussian wave packets of radiation
numerically and found analogous non-linear behaviour to the scalar case.
They also obtained an exact self-similar
critical solution which is qualitatively similar to the discretely self-similar one in the scalar case.  In particular, they found that the critical solution is an intermediate attractor:  as the
critical point is approached, the evolution of the fluid and gravitational field develops a
self-similar region (given by the exact critical solution) near the centre of collapse.
However, only a precisely critical model is described by this solution everywhere.
Maison (1996) extended Evans and Coleman's work to the more general $p = \alpha \rho$ case.  By
considering spherically symmetric non-self-similar perturbations to the critical solution, he
managed to explain the scaling behaviour analytically.  Niemeyer \& Jedamzik (1999) and Shibata \& Sasaki (1999) -- and more recently Musco et al. (2005) -- have extended the Evans-Coleman analysis to models which are asymptotically Freidmann rather than asympotically Minkowski. They find very similar behaviour and, as discussed in Section 6.4,  this has important implications for the formation of primordial black holes.

The numerical simulations of Harada \& Maeda (2001) for the spherical collapse of a $p=\alpha \rho$
fluid also demonstrate the existence of critical phenomena, although the critical solution is
attained only when the initial data are fine-tuned.  This result is supported by their mode analysis of
the self-similar solutions.  Maeda \& Harada (2001) and Harada et al.  (2003) have also studied critical phenomena in Newtonian gravity. In this context, numerical simulations of
the spherical collapse of isothermal gas show that critical behavior at the
threshold of gravitational instability leads to core formation. For a given initial
density profile, they find an associated critical temperature.  For the exact critical
temperature, the collapse converges to the self-similar form associated with the first member of Hunter's family of
self-similar solutions.  In the supercritical case, the collapse first approaches this
solution but then converges to one of the self-similar Larson-Penston solutions. In
the subcritical case, the gas bounces and disperses to infinity.  These critical properties are
similar to those observed in the collapse of a radiation fluid in general relativity.

To identify the global features of
the critical solution, we need to consider the full family of spherically symmetric similarity
solutions discussed in the previous section.  If one confines attention to solutions which are
analytic at the the sonic point, then Carr et al.  (2000) have shown numerically that, for all
$\alpha$, the critical solution starts from a regular centre, passes through a sonic point and
enters the spatially self-similar region.  For $0<\alpha<0.28$, the critical solution is of the
asymptotically quasi-static kind:  it passes through the spatially self-similar region and enters a
second timelike self-similar region. It eventually reaches another sonic point, which is generally
irregular.  However, this does not invalidate the solution as being the critical one, since the
solution describing the inner collapsing region is usually matched to an asymptotically flat
exterior geometry sufficiently far from the centre.  For $0.28<\alpha<0.89$, the critical solution
belongs to the asymptotically Minkowski class associated with finite $z$.  For the limiting
case $\alpha\approx0.28$, the critical solution is also asymptotically Minkowski but of the kind
associated with infinite $z$.  For $0.89<\alpha<1$, the nature of the critical solution is unclear.

If $\alpha=1$,
corresponding to a ``stiff'' fluid with $p=\rho$, the similarity equations are singular and so the
analysis is not so straightforward.  However, it is well known that a stiff fluid is equivalent to a massless scalar field, providing its gradient is everywhere timelike. Spherically symmetric
self-similar spacetimes with a massless scalar field have been investigated by a number of
authors (Brady 1995, Koike et al.  1995, Hod \& Piran 1997, Frolov 1997).
The self-similar
solution of Roberts (1989) is also relevant in this context.  This describes the implosion of
scalar radiation from past null infinity.  The solution is described by a single parameter:  it
collapses to a black hole when this parameter is positive, disperses to future null infinity -
leaving behind Minkowski space - when it is negative, and exhibits a null singularity when it is
zero.  Although this is reminiscent of the usual critical behaviour, the critical solution is
not an intermediate attractor since nearby solutions do not evolve towards it. Self-similar collapse of scalar fields in higher dimensions has also been studied (Frolov 1999a, 1999b).

None of the self-similar critical solutions are stable and, in this sense, they do not satisfy the similarity hypothesis.  However, they are singled out by having only one instability mode and, in this sense, they might be regarded as being as stable as possible (Maison 1996). The breaking of self-similarity in the critical collapse of a scalar field has been studied in detail by Frolov (1997, 2000). He shows that the generic growing perturbation departs from the Roberts solution in a universal way:  the initial continuous self-similarity of the background is broken into a discrete self-similarity, with exactly the echoing period found by Choptuik. On the other hand, an analysis of non-spherical perturbations of the Roberts solutions shows that there are no growing non-spherical modes (Frolov 1998).

\subsection{Stability analysis of spherically symmetric self-similar models}Presumably a necessary (but not sufficient) condition for spherically symmetric solutions to tend to self-similar form is that self-similar solutions (or at least some subset of them) be stable to non-self-similar spherically symmetric perturbations. There are several contexts in which this problem has already been studied. For example, the stability of particular self-similar solutions is already implicit in the stability analysis of the critical solutions discussed at the end of Section 6.2. We have seen that these solutions are necessarily self-similar (asymptotically quasi-static or Minkowski, depending on the value of $\alpha$) and they are always unstable to a single mode.
Another example -- in both the Newtonian and relativistic contexts -- is the well-known ``kink'' instability: solutions which are non-analytic at the sonic pointeither develop shocks or are driven towards analytic ones (Whitworth and Summers 1985, Ori and Piran 1990). This raises the question of whether the asymptotic self-similar solutions postulated by the similarity hypothesis are analytic or contain shocks.

Harada (2001) has extended this type of analysis by studying the stability at the sonic point for more general self-similar solutions containing a $p=\alpha \rho$ fluid.  (The different types of behaviour at the sonic point were discussed in Section 5.2.) For the collapsing case, all
primary-direction nodal-point solutions are unstable, while all secondary-direction nodal-point
solutions and saddle-point ones are stable.  The situation is reversed in expansion.  This means
that expanding flat Friedmann solutions with $1/3 \le \alpha < 1$ and collapsing ones with $0<
\alpha \le 1/3$ are unstable. The Larson-Penston solution is stable for $0<\alpha<0.036$ and unstable for $\alpha>0.036$; the Evans-Coleman solution is stable for $0<\alpha<0.89$ and unstable
for $1>\alpha>0.89$. The last application suggests that the Evans-Coleman solution for $1>\alpha>0.89$
is not the critical one because it has two unstable modes.

Recently there have been several further developments in this area. Harada and Maeda (2004) have analysed the scalar field and stiff fluid cases ($\alpha =1$) and again find that a wide class of the self-similar solutions are unstable against kink mode perturbations. In this case, the sonic point coincides with the particle horizon for cosmological solutions. In particular, the Evans-Coleman stiff-fluid solution is unstable, so again this cannot be the critical solution. The self-similar scalar-field solution found numerically by Brady et al. (2002) is also unstable.  Maeda \& Harada (2005) have now extended this analysis to include the case of a scalar field with a potential. The analysis of the kink effect by Wang \& Wu (2005) includes the behaviour of the solutions at spatial infinity.

All of these studies focus on the stability of {\it transonic} self-similar solutions and focus on the behaviour at the sonic point.  As a first step to studying the problem in a more general context, Carr and Coley (2005; CC)have investigated the stability of spherically symmetricsimilarity solutions within the more general class of spherically symmetricsolutions. Following Cahill and Taub (1971), they express all functions interms of the similarity variable $z=r/t$ and the radial coordinate $r$ andregard these as independent variables rather than $r$ and $t$. They alsoassume that the perturbations in any quantity $x$ can beexpressed in the form$x(z,r) = x_o (z)[1+x_1 (r)]$,       where a subscript 0 indicates the form of the function in the exactself-similar case and a subscript 1 indicates the fractional perturbationin that function (taken to be small; i.e., $x_1<<1$). Theperturbation equations for $x_1(r)$ can then be expressed as second orderdifferential equations in $r$ and CC test whether a particularsimilarity solution is stable by examining whether theperturbation terms grow or decay at large values of $r$. For example, they find that
the asymptotically flat Friedmann solutions are stable providing $\alpha > -1/3$ and this directly relates to the issue of whether density perturbations evolve to self-similar form.  They also find that the asymptotically Kantowski-Sachs solutions are unstable for $1>\alpha>-1/3$ but stable for $-1<\alpha<-1/3$ (probably the only physically realistic case). This relates to the formation of bubbles in an inflationary scenario (Wesson 1986) and hence tothe stability of the  inflationary phase. However, it should be cautioned that the form of the initial perturbation assumed by CC may not be sufficiently general. This is because the function $x_1$ could depend on $z$ as well as $r$ (this indeed is the case in the analyses of Harada and colleagues), so CC are currently extending their calculations.

\subsection{Can black holes grow as fast as the universe?}
Self-similar asymptotically Friedmann solutions containing black holeswere originally studied because there wasinterest in whether primordial black holes (PBHs) could grow at the same rate as the particlehorizon. A simple Newtonian analysis suggested that this could happen if the black hole had a size comparable to the particle horizon at formation (Zeldovich and Novikov 1967), and this might indeed be expected. Carr and Hawking (1974) showed that self-similar solutions containing a black holeexist for radiation ($\alpha=1/3$) and dust ($\alpha=0$) but only if the universe is asymptotically rather than exactly Friedmann. There is no solution in which a black hole is attached to an exact Friedmann background via a sound-wave, which means that black holes formed through purelylocal processesin the early Universe cannot grow as fast as the particle horizon. (In fact, even the asymptotically Friedmann black hole solutions have no sound-waves because the fluid flow is everywhere supersonic.) Carr (1976) and Bicknell and Henriksen (1978a) then extended this result toa general $0<\alpha<1$ fluid. Lin et al. (1976) claimed that there {\it is}a similarity solution in an exact Friedmann universe for the special case ofa stiff fluid ($\alpha=1$) but Bicknell and Henriksen (1978b) showed thatthis requires the inflowing material to turn into a null fluid at the eventhorizon, which is presumably unphysical. These results show that the similarity hypothesis is certainly not always applicable.

This conclusion is supported by numerical calculations. If the formation and evolution of a PBH in a general fluid universe
with a local perturbation is simulated numerically, without assuming
self-similarity, it is found that the PBH
soon becomes much smaller than the
cosmological horizon and this excludes self-similar
growth (Nadezhin et al. 1978, Novikov and Polnarev 1980). The study of critical phenomena in a Friedmann background (Niemeyer and Jedamzik 1999, Shibata and Sasaki 1999, Musco et al. 2005) also gives no evidence for the self-similar growth of PBHs formed from primordial density perturbations. Indeed the nature of the critical solution implies that growth can be self-similar only in the limit that the PBH mass goes to zero.

It is interesting to consider whether this result also applies for a black hole in a universe whose density is dominated by a scalar field.
In particular, the question of whether PBHs can grow as fast as the universes in the quintessence scenario
has attracted attention.  If the scalar field is massless and there is no scalar potential, then we have seen that it is equivalent to
a stiff fluid, so the fact that  there is no self-similar solution in the stiff fluid case suggests that accretion is also limited in the scalar field case. Indeed, recent numerical calculations by Harada and Carr (2005a, 2005b) support this conclusion. Nevertheless, a variant of the original Newtonian argument
for self-similar growth has recently been applied in the quintessence scenario to argue that PBHs could grow enough to form the supermassive black holes found in galactic nuclei (Bean and Magueijo 2003). A generalization of this analysis (Custodio and Horvath 2005), not necessarily involving self-similarity but still based on the Newtonian analysis, has also claimed there could be appreciable quintessence accretion in some circumstances. Both these studies incorporate a scalar potential, so the stiff fluid analysis is not necessarily applicable, but the invocation of large accretion is clearly questionable (cf. Harada et al. 2005).

\section{Alternative Theories and the Early Universe}

In studies of the early Universe, new physics -- such as
string theory -- is necessary close the Planck time. Although
a complete fundamental theory is not presently known, the
phenomenological consequences can be understood by studying an
effective low-energy theory, which leads to the introduction of
additional (usually scalar) fields in
the high-curvature regime (e.g., the dilaton of string theory or moduli fields). Indeed, scalar fields are
believed to be pervasive in all fundamental theories
of physics applicable in the early Universe. Similar considerations arise if physics near the big bang  involves a scalar-tensor theory of gravity, since this is conformally related to general relativity plus a scalar field.

A massless scalar field has an effective equation of state $\alpha =1$ close
to the initial singularity. This already has important implications, since -- as  discussed in Section 3.1 --
a generic inhomogeneous cosmology in general relativity with a stiff fluid or scalar field
tends to a velocity-dominated solution at early times (Andersson and Rendall 2001).
There are also a number of cosmological models of current physical interest -- such as the ekpyrotic
or cyclic cosmological models -- which
have an effective equation of state parameter  $\alpha \ge 1$. In such models, it has been
shown (Erickson et al., 2004) that chaotic Mixmaster
oscillations due to anisotropy and curvature are suppressed and the
contraction is spatially homogeneous and isotropic.

Recent developments have motivated the idea that
gravity may be a truly higher-dimensional theory, becoming
effectively 4-dimensional at lower energies. In particular, there is
currently great interest in models inspired by string theory, in
which the matter fields are confined to a ``brane'' embedded in
higher dimensions (Rubakov and Shaposhnikov 1985). This has led to
higher-dimensional cosmological models (Randall and Sundrum, 1999), in which a $\rho^2$ term dominates the Friedmann equation on the brane at early times. This is associated with an effective equation of state with $\alpha  > 1$.
It has been found that an isotropic singularity is a
past-attractor in all orthogonal Bianchi brane-world cosmological models
(Coley, 2002; van den Hoogen et al., 2003) and also in a class of spatially
inhomogeneous brane-world models
(Coley, 2002; Coley et al., 2004).

The past asymptotic decay rates have been calculated analytically close to the
initial singularity in general $G_{0}$ spatially inhomogeneous models with
an effective equation of state $\alpha
\ge 1$  by  Coley and Lim (2005), confirming
the above results. They have also studied a special class of $G_{2}$ cosmological models numerically. In particular, a subset of the Jacobs  solutions for
$\alpha =1$ and  the flat Friedmann-Lema\^{\i}tre solution for
$\alpha  >1$  are locally stable towards the past. In all cases, the singularity is self- similar. However,
it may be necessary to investigate the robustness of these results in the presense of
additional fields. For example, p-form fields in
string or M-theory can lead
to oscillatory behaviour close to cosmological singularity (Damour et al. 2002).

There is evidence that self-similarmodels play an important dynamical role in alternativetheories of gravity, such as scalar-tensor theories orstring or $M$ theory (Coley 2000).
In particular, it is of interest to study the cosmological
implications of $M$-theory, whose moduli space contains all five
anomaly-free ten-dimensional superstring and eleven-dimensional
supergravity theories. In Coley
(2003) the asymptotic dynamical evolution of brane-world models close to the initial singularity was
discussed. It was shown that  the initial
singularity is generically isotropic in spatially homogeneous cosmological
models and it is plausible that this is also true in inhomogeneous brane-world
models. Therefore, an isotropic singularity (Goode and Wainwright
1982, Goode et al. 1992) is a (local) past-attractor. This means that the evolution near the
cosmological initial singularity is approximated by the flat FRW
model, so the universe is asymptotically self-similar in
these models.

Homogeneous plane-wave spacetimes are plane-fronted waves
(i.e., pp-waves with additional ``planar'' symmetry along the
wavefronts). There are exact solutions to string
theory of this kind if supported by an appropriate antisymmetric field and
dilaton (Amati and Klimcek 1988, Horowitz and Steif 1990) and they can be quantized
(Gibbons 1999). The
maximally supersymmetric BFHP (Blau et al. 2002) plane-wave
solution of type-IIB string theory was also found to be exactly
solvable.  Any general relativistic solution admits plane-wave backgrounds in
the Penrose limit (Penrose 1976).
This was extended to supergravity solutions
by Gueven (1987). The super-pp-wave BFHP
solution arises as the Penrose-Gueven limit of  $AdS_5 \times
S^5$, and this gives rise to a novel explicit form of the AdS/CFT
correspondence (Maldacena 1998, Berenstein et al. 2002). Subsequently, the connection between
supergravity and gauge theories has been further explored
and the Penrose limit has been studied for
hints on how to  quantize strings on more general
backgrounds. The non-trivial homogeneous plane-wave spacetimes are precisely the
one that occur as the Penrose limits of solutions of
supergravity. We note that all of these spacetimes admit a
homothetic vector (Coley et al. 2004) and are consequently
self-similar.
\section{Discussion}

It is clear that self-similar solutions play a crucial role in a wide range of relativistic and
Newtonian problems. This applies in both the spatially homogenous and spherically symmetric
contexts. So although the literature cited in Sections 2-4 and Sections 5-6 of this paper is fairly
disjoint, there are striking parallels between the two areas.  In both contexts a dynamical systems
analysis reveals that self-similar solutions can act as attractors and there are many examples in
which a self-similar solution is an asymptote at either late, early or intermediate times.  It is
therefore clear that the ``similarity hypothesis", as we have termed it, applies in many
important situations.
In fact, the status of the similarity hypothesis in the spatially homogeneous case is uncontroversial, since one can show explicitly which solutions are asymptotically self-similar.
Its status in the spherically symmetric case is less clear from a mathematical perspective, although most of the important physical applications arise in this context.

Of course, the similarity hypothesis does not always apply, since there are certainly some situations in which
models are {\it not} expected to be asympotically self-similar.  For example, in the context of the
spatially homogeneous models, we have seen that Bianchi types VIII and IX exhibit
(non-self-similar) oscillatory Mixmaster behaviour as one goes back to the initial singularity.
Also the late-time evolution of non-tilted spatially homogeneous models is dominated by the Weyl
curvature for Bianchi VII$_o$ and Bianchi VIII models if $0<\alpha<1$, so again there is no
asymptotic self-similarity.  Likewise, in the spherically symmetric situation, we
have seen that dust solutions cannot generally satisfy the similarity hypothesis, because their
evolution is prescribed by the ``energy function" and they can only be self-similarity if this is
constant.

The challenge therefore is to identify the conditions under which the similarity hypothesis
applies.  One possibility in the spherically symmetric context is that we may need shocks or pressure
to force asymptotic similarity.  Indeed the existence of the ``kink" instability suggests that self-similar solutions with shocks may sometimes be the generic asymptotic state. On the other hand, this alone does not suffice, since we have seen that there
are no self-similar solutions in which a black hole is attached to an exact exterior Friedmann
solution at a sonic point.  Although there are self-similar solutions if the exterior is only
asymptotically Friedmann, these do not contain a sonic point and -- as in the dust case -- they
require very special initial conditions.

It should be stressed that nearly all of our considerations have been confined to a single
perfect fluid.  This is a significant restriction since multi-fluids sometimes exhibit
important physical features which would not appear in the single fluid case.  For example, the
two-stream instability that leads to mass inflation in gravitational collapse (Poisson
1990) is suppressed if a single fluid is assumed.  Since the existence of self-similar
mass-inflation solutions has been demonstrated in the two-fluid case (charged baryons and dark
matter) by Hamilton \& Pollack (2005a), self-similar solutions which do not admit mass
inflation may be a poor guide to reality.  The assumption of a perfect fluid may also be
physically restrictive, since self-similar mass-inflation solutions also exist with a single
fluid if one includes conductivity (Hamilton \& Pollack 2005b).

Several studies go beyond the single perfect fluid situation.  For example, spherically symmetric self-similar solutions containing a perfect fluid and a scalar field have been studied by Coley \& Goliath (2000) and those containing two scalar fields have been studied by Coley \& Taylor (2001) and Coley \& He (2002a, 2002b). Spherically symmetric self-similar solutions with mixed fluids were also discussed briefly by Carr \& Coley (2000). However, there is not yet a complete classification of such solutions, so the status of the similarity hypothesis in this context remains unclear.

Finally, we have seen that the similarity hypothesis may also have
important applications in a wider context than general relativity, including alternative theories of gravity (e.g.,
string and M-theories) and higher- dimensional cosmological scenarios (e.g., ekpyrotic and brane models).
In all of these cases, the asymptotic states of models are self-similar, which may be of
physical significance.

%%%%Another point is that, although we have classified all the spherically symmetric self-similar solutions, it is not clear which of these can serve as asymptotes. Black holes in an exact  Friedmann background may not be described by a similarity solution but perhaps voids or overdense regions can be.
%we have seen that all (non-tilting)
%perfect fluid models with a linear equation of state parameter are
%asymptotically self-similar except for (i) the non-stiff VIII/IX
%models to the past and (ii) the VII$_0$ and VIII/IX models to the future.

%NO REFS PRINTED OUT\end{document}\section*{Acknowledgements}This work was supported, in part, by NSERC of Canada. We thank A. Hamilton and T. Harada for useful comments on a first draft of this paper. \subsection*{REFERENCES}

\begin{enumerate}

\item[] D. Amati and C. Klimcik, 1988, Phys. Lett. {\bf B210}, 92.

\item[] D. Amati and C. Klimcik, 1989, Phys. Lett.  {\bf B 219}, 443.

\item[] L. Andersson and A. D. Rendall, 2001,
Commun. Math. Phys. {\bf 218}, 479 [gr-qc/0001047].

\item[] G. I. Barenblatt, 1952, Prikl. Mat. Mekh. {\bf 16}, 67.
\item[] G. I. Barenblatt and Ya B. Zeldovich, 1972, Ann. Rev. Fluid Mech. {\bf 4}, 285.

\item[]  J. D. Barrow and F. J. Tipler, 1986, {\it The AnthropicCosmological Principle} (Oxford University Press).

\item[]
J. D. Barrow and S. Hervik, 2003,
 Class. Quant. Grav. {\bf 20},  2841  [gr-qc/0304050].

\item[] J. D. Barrow and C. Tsagas, 2005, Class. Quant. Grav. {\bf 22}  825
[gr-qc/0411070].

 

\item[] R. Bean and J. Magueijo, 2002, Phys. Rev. D {\bf 66}, 063505.
\item[]  V A.  Belinskii, I.  M.  Khalatnikov and E.  M.  Lifshitz, 1970, Adv.Phys.  {\bf 19} 525.

\item[] V.  A.  Belinskii, I.  M.  Khalatnikov and E.  M.  Lifshitz, 1971, Sov. Phys.  Usp. {\bf 13}, 745.\item[] V.A.  Belinskii and I.M.  Khalatnikov, 1972, \v Z.  \`Eksper.  Teoret.Fiz.  {\bf63}, 1121. \item[] V.  A.  Belinskii, I.  M.  Khalatnikov and E.  M.  Lifshitz, 1982, Adv.  Phys.  {\bf31}, 639.\item[] P.  M.  Benoit and A.  A.  Coley, 1998, Class.  Quantum Grav.  {\bf 15}, 2 397.

\item[] D. Berenstein,  J. Maldacena and H. Nastase, 2002,
JHEP {\bf 0204}, 013.

\item[] B.K.  Berger and D.  Garfinkle, 1998, Phys.  Rev.  D {\bf57}  4767 [gr-qc/971012].

 \item[] B.  K.  Berger and V.  Moncrief, 1983, Phys.  Rev.  D {\bf48}, 4676.

 \item[] B.  K.  Berger and V.  Moncrief, 1998, Phys.  Rev.  D {\bf57} 7235 [gr-qc/9801078].

\item[] E.  Bertschinger, 1985, Ap.  J.  {\bf 268}, 17.

\item[] E.  Bertschinger and P.  N.
Watts, 1984, Ap.  J.  {\bf 328}, 23.

\item[] G.  V.  Bicknell and R.  N.  Henriksen, 1978a, Ap.
J.  {\bf 219}, 1043.

 \item[] G.  V.  Bicknell and R.  N.  Henriksen, 1978b, Ap.  J.  {\bf 225},
237.

\item[] G.  V.  Bicknell and R.  N.  Henriksen, 1979, Ap.  J.  {\bf 232}, 670.

\item[]  A.P. Billyard, A.A. Coley
and J.E. Lidsey, 1999, J. Math. Phys.   {\bf 41}, 6277.

\item[]  A.P. Billyard, A.A. Coley
and J.E. Lidsey, 2000,
Class. Quant. Grav.  {\bf 17}, 453.

\item[] A.
P.  Billyard, A.  A.  Coley, R.  J.  van den Hoogen, J.  Iba\~nez and I.  Olasagasti, 1999,
Class.  Q.  Grav.  {\bf16}, 4035
[gr-qc/9907 053].

\item[]  A.P. Billyard, A.A. Coley,
J.E. Lidsey and U.S Nilsson, 2000,
Phys. Rev. D  {\bf 61}, 043504.

\item[]  P. Bizon and A. Wasserman, 2002, Class. Quant. Grav.  {\bf 19}, 3309.

\item[] M.  Blau,  J. Figueroa-O'Farrill, C. Hull and G. Papadopoulos (BFHP), 2002, JHEP {\bf 0201}, 047; see also
hep-th/0202111.

 

 \item[] H.  Bondi, 1947, MNRAS {\bf 107}, 410.

 \item[] W.  B.  Bonnor, 1956,
Z.  Astrophys.  {\bf 39}, 143.

\item[] W. B. Bonnor, 1974, MNRAS {\bf 167}, 55.

\item[] P. R. Brady, 1995, Phys. Rev. D {\bf 51}, 4168.

\item[] P. R. Brady, 1999, Prog. Theor. Phys. Supp. {\bf 136}, 29.

\item[] P. R. Brady et al., 2002, Class. Quantum. Grav. {\bf 19}, 6359.

\item[] M. Bruni, S. Matarrase and O. Pantano, 1995, Phys. Rev. Lett. {\bf74}, 1916 [astro-ph/94074].

 

\item[] A. H. Cahill and M. E. Taub, 1971, Comm. Math. Phys. {\bf 21}, 1.

\item[] R.R. Caldwell, R. Dave and P.J. Steinhardt, 1998,  Phys. Rev. Lett. {\bf 80}, 1582.\item[]  B. J. Carr, 1993, preprint prepared for but omitted from {\it The Origin of Structure in the Universe}, ed. E. Gunzig and P. Nardone (Kluwer).

\item[] B. J. Carr, 2000, Phys. Rev. D. {\bf 62}, 044022.
\item[]  B. J. Carr and A. A. Coley, 1999, Class
Quantum Grav. {\bf  16} R31.
\item[] B. J. Carr and A.A. Coley, 2000, Phys. Rev. D {\bf 62} 044023.

\item[]  B. J. Carr and A.A. Coley (CC), 2005, in preparation.

\item[] B. J. Carr and C. Gundlach, 2003, Phys. Rev. D. {\bf 67}, 024035.
\item[] B. J. Carr and S. W. Hawking, 1974, MNRAS {\bf 168}, 39.\item[] B. J. Carr and A. Koutras, 1992, Ap. J. {\bf 405}, 34.

\item[] B. J.  Carr and A. Yahil, 1990, Ap. J. {\bf 360}, 330.

\item[]  B. J. Carr, A. A. Coley, M. Goliath,  U.S. Nilsson and C. Uggla, 2000,
  Phys.  Rev.  D. {\bf 61}, 081502.

\item[]  B. J. Carr and A. A. Coley, M. Goliath,  U.S. Nilsson and C. Uggla, 2001,
Class. Quant. Grav.   {\bf 18}, 303.

 \item[] B.  Carter and R.  N.  Henriksen, 1989, Ann.  Physique Supp.
{\bf 14}, 47.

 \item[] B.  Carter and R.  N.  Henriksen, 1991, J.  Math.  Phys.  {\bf32}, 2580.\item[]  M. W. Choptuik, 1993, Phys. Rev. Lett. {\bf 70}, 9.\item[]  M. W. Choptuik, 1994, in {\it Deterministic Chaos in GeneralRelativity}, ed. D. Hobill et al. (Plenum, New York).

\item[] D. Christodoulou, 1984, Commun. Math. Phys. {\bf 93}, 171.

\item[] D. Christodoulou, 1994, Ann. Math.  {\bf 140}, 607.

\item[] D. Christodoulou, 1999, Commun. Math. Phys. {\bf 149}, 183.
 \item[] C.A.  Clarkson, A.A.  Coley and S.D. Quinlan, 2001,
Phys. Rev. D  {\bf 64}, 122003  [astro-ph/0108268].
\item[] A. A. Coley, 1997, Class. Quantum Grav. {\bf 14}, 87.\item[] A.A. Coley, 2000,  {\it Dynamical systems in cosmology}, in {\it Recent Developments in Gravitation, Proceedingsof the Spanish Relativity Meeting ERE-99}, ed.  J. Iba\~nez, pp 13-44  [gr-qc/9910074].
 \item[] A. A. Coley, 2002,
Class. Quant. Grav. {\bf 19}, L45.
[hep-th/0110117].
\item[] A. A. Coley, 2002, Phys.
Rev. D. {\bf 66}, 023512.

\item[] A.A. Coley, 2003,
{\it Dynamical systems and cosmology},
Kluwer Academic Publishers.

\item[] A.A. Coley and M. Goliath, 2000,
Class. Quant. Grav.  {\bf 17}, 2557 [gr-qc/0003080].

\item[] A. Coley and Y. He, 2002a,
Class. Quant. Grav.  {\bf 19}, 3901.

\item[] A. Coley and Y. He, 2002b,
Gen. Rel. Grav.  {\bf 35}, 707

\item[] A. Coley and  S. Hervik, 2004,
Class. Quant. Grav. {\bf21}, 4193 [gr-qc/0406120].

\item[] A. Coley and  S. Hervik, 2005,
Class. Quant. Grav. {\bf 22}, 579  [gr-qc/0409100].

\item[] A. A. Coley and R. J. van den Hoogen, 1994, J. Math. Phys. {\bf35}, 411.

\item[]
A. A. Coley and R. J. van den Hoogen, 1995,  Class. Quant. Grav.{\bf 12} 2045.

\item[] A. A. Coley and W. C. Lim, 2005, Class. Quant. Grav. {\bf22} 3073
[gr-qc/0506097].

 
\item[] A. A. Coley and T. D. Taylor, 2001,
Class. Quant. Grav.  {\bf 18}, 4213.

\item[] A. A. Coley and J. Wainwright, 1992, Class. Quantum Grav. {\bf 9},651.
\item[] A. A. Coley, Y. He and W. C. Lim,  2004,
Class. Quant. Grav. {\bf 21},  1311 [gr-qc/0312075].

\item[] A.A. Coley, R.Milson, V. Pravda and
 A. Pravdova, 2004,
 Class. Quant. Grav. {\bf 21}5519 [gr-qc/041007].

 

\item[] C. B. Collins, 1971, Comm. Math. Phys. {\bf 23}, 13.\item[] C. B. Collins, 1977, J. Math. Phys. {\bf 18}, 2116.\item[] B. C. Collins, 1985, J. Math. Phys. {\bf 26}, 2268.\item[] C. B. Collins and S. W. Hawking, 1973, Ap. J. {\bf 180}, 317.\item[] C. B. Collins and J. Stewart, 1971, MNRAS {\bf 153}, 419.\item[] E.J. Copeland, A.R. Liddle and D. Wands, 1998, Phys. Rev. D {\bf 57}, 4686.

\item[] P. S. Custodio and J. E. Horvath, 2005,
Int. J. Mod. Phys. {\bf 14}, 257 [gr-qc/0502118].

\item[] T. Damour, M. Henneaux, A. D. Rendall, M. Weaver, 2002, Ann. H. Poincare {\bf 3}, 1049.

\item[] D. M. Eardley and L. Smarr, 1979, Phys. Rev. D {\bf 19}, 2239.
\item[] G. Efstathiou et al., 1990, MNRAS {\bf 247}, 10.

\item[] G.F. R. Ellis, 1988,  Class. Quantum Grav. {\bf 5}, 891.

\item[] J. K. Erickson, D. H. Wesley, P. J. Steinhardt and N. Turok, 2004,
Phys. Rev. D {\bf69}, 063514

\item[] C. R. Evans and J. S. Coleman, 1994, Phys. Rev. Lett. {\bf 72},1782.\item[] J. A. Fillmore and P. Goldreich, 1984, Ap. J. {\bf 281}, 1.\item[] T. Foglizzo and R. N. Henriksen, 1993, Phys. Rev. D. {\bf48}, 4645.

\item[] A. V. Frolov, 1997, Phys. Rev. D {\bf 56}, 6433.

\item[] A. V. Frolov, 1999a, Phys. Rev. D {\bf 59}, 104011.

\item[] A. V. Frolov, 1999b, Class. Quantum Grav. {\bf 16}, 407.

\item[] A. V. Frolov, 2000, Phys. Rev. D {\bf 61}, 084006.

%\item[] J.  Frauendiener and B.  G.  Schmidt, 1992.

\item[] C.  S.  Frenk et al., 1988, Ap.  J.  {\bf 351}, 10.

\item[] P.  M.  Garnavich et al., 1998,
Ap.  J.  {\bf 509}, 74.

\item[] G.W. Gibbons,
1999,
 Class. Quantum Grav. {\bf 16}, L71.

 \item[] M.  Goliath, U.  S.  Nilsson
and C.  Uggla, 1998, Class.  Quant.  Grav.  {\bf 15}, 167.

\item[] S.  W.  Goode and J.
Wainwright, 1982, Class.  Quantum Grav.  {\bf 2}, 99.

\item[]  S.W. Goode, A.A. Coley and J. Wainwright, 1992,
Class. Quant. Grav.  {\bf 9}, 445.

\item[] R.~Gueven, 1987, Phys.\ Lett.\ B
{\bf 191}, 275.

%\item[] C. Gundlach, 1999, gr-qc/9906124.

\item[] C. Gundlach, 1999, Living Review {\bf 2}, 4.

\item[] C. Gundlach, 2003, Physics Reports {\bf 376} 339 [gr-qc/0210101].

 

%\item[] C. Gundlach and J. Martin-Garcia, 2000, %gr-qc/9906068.\item[] J. E.  Gunn, 1977, Ap. J. {\bf 218}, 592.\item[] J. E. Gunn and J. R. Gott, 1972, Ap. J. {\bf 176}, 1.

\item[] A.J.S. Hamilton and S.E. Pollack, 2005a, Phys. Rev. {\bf D71}, 084031.

\item[] A.J.S. Hamilton and S.E. Pollack, 2005a, Phys. Rev. {\bf D71}, 084032.

\item[] M. J. Hancock, D. The, J. Wainwright, 2003,
Class. Quant. Grav. {\bf 20} 1757.

\item[] T. Harada, 2001, Class. Quant. Grav. {\bf 18}, 4549.

\item[] T. Harada and H. Maeda, 2001, Phys. Rev. D {\bf 63}, 084022.

\item[] T. Harada and H. Maeda, 2004, Class. Quant. Grav. {\bf 21}, 371.

\item[] T. Harada and B.J.Carr, 2005a, Phys. Rev. D.  {\bf 71}, 104009.

\item[] T. Harada and B.J.Carr, 2005b, Phys. Rev. D. {\bf 71}, 104010.

\item[] T. Harada, H. Maeda and B. Semelin, 2003, Phys. Rev. D {\bf 67}, 084003.

\item[] T. Harada, H. Maeda and B. Carr, 2005, in preparation.

\item[] M. A. Hausman et al., 1983, Ap. J. {\bf 270}, 351.%\item[] C. Hellaby, 19XX, J. Math. Phys. {\bf 37}, 2892.
\item[] R.  N.  Henriksen, 1989, MNRAS, {\bf 240}, 917.
\item[] R. N. Henriksen and K. Patel, 1991, Gen. Rel. Grav. {\bf 23}, 527.
\item[] R.  N.  Henriksen and P. S. Wesson, 1978, Astrophys. Sp. Sci.  {\bf 53}, 429 \& 445.

\item[]  S.~Hervik, 2004, Class. Quant. Grav. {\bf 21}, 2301.

\item[] S. Hervik and A. Coley, 2005,  Class. Quant. Grav. ???  [gr-qc/0505108].

\item[] S. Hervik, R. van den Hoogen and  A. Coley, 2005,  Class. Quant. Grav. {\bf 22},
607 [gr-qc/0409106].

\item[] C. G. Hewitt and J. Wainwright, 1993, Class. Quantum Grav. {\bf10}, 99.

\item[] C.G.~Hewitt, R.~Bridson and J.~Wainwright, 2001,
Gen. Rel. Grav. {\bf 33}, 65.

%\item[] C. G. Hewitt and J. Wainwright, 1990, Class. Quantum Grav. {\bf%7}, 2295.%\item[] C. G. Hewitt and J. Wainwright, 1992, Phys. Rev. D {\bf 46},%4242.
\item[]
C. G. Hewitt, J. T. Horwood and J. Wainwright, 2003,
Class. Quant. Grav. {\bf 20}, 1743.

\item[] C. G. Hewitt, J. Wainwright and S. W. Goode, 1988, Class. Quantum Grav. {\bf 5}, 1313.\item[] C. G. Hewitt, J. Wainwright and M. Glaum, 1991, Class. QuantumGrav. {\bf 8}, 1505.
\item[] D.~Hobill, A.~Burd and A.A.~Coley (ed), 1994,
{\it Deterministic Chaos in General Relativity},
Plenum Press.
\item[] Y. Hoffman and J. Shaham, 1985, Ap. J. {\bf 297}, 16.

\item[] S. Hod and T. Piran, 1997, Phys. Rev. D. {\bf 55}, R440.
\item[] G.T. Horowitz and A.R. Steif, 1990, Phys. Rev. Lett. {\bf
64}, 260; {\bf 42}, 1950.

\item[] J.T.~Horwood and J.~Wainwright, 2004,
Gen. Rel. Grav. {\bf 36}, 799.

\item[] L. Hsu and J. Wainwright, 1986, Class. Quantum Grav. {\bf 3},1105.
\item[] C. Hunter, 1977, Ap.J. {\bf 218}, 8348.

\item[] C. Hunter, 1986, MNRAS {\bf 223}, 391.
\item[] S. Ikeuchi, K. Tomisaka and J. P. Ostriker, 1983, Ap. J. {\bf265}, 583.

\item[] K. Jedamzik and J. C. Niemeyer, 1999, Phys. Rev. D {\bf 59}, 124014.
\item[]  R. Kantowski and R. Sachs, 1966, J. Math. Phys. {\bf 7}, 443
\item[] S.  Kichenassamy and A.  D. Rendall, 1998, Class. Quant. Grav.    {\bf15   }, 1339.

\item[] T. Koike. T. Hara and S. Adachi, 1995, Phys. Rev. Lett.  {\bf 74}, 5170.

\item[] D. A. Konkowski and T. M. Helliwell, 1996, Phys. Rev. D {\bf 54}, 7898.

\item[] D. Kramer, H. Stephani, M. A. H. MacCallum and E. Herlt, 1980, {\it Exact Solutions of Einstein's Field Equations}(Cambridge University Press, Cambridge).
\item[] A. Krasinzki, 1997, {\it Physics in an Inhomogeneous Universe} (CambridgeUniversity Press, Cambridge). \item[] K. Lake, 1992, Phys. Rev. Lett. {\bf 68}, 3129.

\item[] K. Lake and T. Zannias, 1990, Phys. Rev. D {\bf 41}, 3866.

\item[] R. B. Larson, 1969, MNRAS {\bf 145}, 271.

\item[] V. G. Leblanc, 1997, Class. Quant. Grav.  {\bf 14}, 2281.

\item[] V. G. Leblanc, 1998, Class. Quant. Grav.  {\bf
15}, 1607.

\item[] E.M.  Lifshitz and I.M.
Khalatnikov, 1963,  Advances in Physis {\bf 12}, 185.

\item[] W.C.~Lim, H.~van Elst, C.~Uggla and J.~Wainwright, 2004,
Phys.  Rev. D {\bf 69}, 103507.

\item[] D. N. C. Lin, B. J. Carr and S. D. M. Fall, 1978, MNRAS {\bf 177}, 151.

\item[]  J.E. Lidsey, D. Wands and E.J. Copeland, 2000, Phys. Rep. {\bf
337}, 343.
\item[] X. Lin and R. M. Wald, 1989, Phys. Rev. D {\bf 40}, 3280.\item[] D. Lynden-Bell, 1967, MNRAS {\bf 136}, 101.

\item[] D. Lynden-Bell and J.P.S. Lemos, 1988, MNRAS {\bf 233}, 197.
\item[] H. Maeda and T. Harada, 2001, Phys. Rev.  {\bf D64},
124024.

\item[] H. Maeda and T. Harada, 2005, Phys. Lett. {\bf B607}, 8.
124024.

\item[] H. Maeda, T, Harada, H. Iguchi and N. Okuyama, 2002a, Phys. Rev. {\bf D66}, 027501.

\item[] H. Maeda, T, Harada, H. Iguchi and N. Okuyama, 2002b, Prog. Theor. Phys. {\bf 108}, 819.

\item[] H. Maeda, T, Harada, H. Iguchi and N. Okuyama, 2003, Prog. Theor. Phys. {\bf 110}, 25.
\item[] D. Maison, 1996, Phs. Lett. B. {\bf 366}, 82.
\item[] J. Maldacena, 1998, Adv. Theor. Math. Phys. {\bf 2},  231.

\item[]  C. W. Misner, 1968, Ap. J. {\bf 151}, 431.\item[] C. W. Misner and H. S. Zapolsky, 1964, Phys. Rev. Lett. {\bf 12},635.

\item[] I. Musco, J. Miller  and L. Rezzolla, 2005,
Class.Quant.Grav.
{\bf 22} 1405.

\item[] D. K. Nadezhin, I. D. Novikov and A. G. Polnarev, 1978, Sov. Astron. {\bf 22}, 129.
\item[] J. C. Niemeyer and K. Jedamzik, 1999, Phys. Rev. D {\bf 59}, 124013.
\item[] U. S. Nilsson and C. Uggla, 1997, Class. Quantum Grav. {\bf 14}  1965.

 \item[] U. S. Nilsson, M. J. Hancock, and
J. Wainwright, 2000, Class.  Quant. Grav. {\bf 17}, 3119  [gr-qc/9912019].

\item[] U. S. Nilsson, C. Uggla and J. Wainwright,
2000, Gen. Rel. Grav.  {\bf 32 } 1319 [gr-qc/9908062].

 
\item[] B. C. Nolan, 2001, Class. Quantum Grav. {\bf 18}, 1651.

\item[] B. C. Nolan and T. J. Waters, 2002, Phys. Rev. Lett. {\bf 66},104012.

\item[] I. D. Novikov and A. G. Polnarev, 1980, Sov. Astron. {\bf 24}, 147.
\item[] K. Olive,1990,
Phys. Rept. {\bf 190}, 307.
 

\item[] A. Ori and T. Piran, 1987, Phys. Rev. Lett. {\bf 59}, 2137.\item[] A. Ori and T. Piran, 1990, Phys. Rev. D. {\bf 42}, 1068.\item[]
 R. Penrose, 1976,
in {\it Differential geometry and relativity}, eds. T. Cahen and
Flato (Reidel).

\item[] M. V. Penston, 1969, MNRAS {\bf 144}, 449.

\item[] S. Perlmutter et al., 1999, Ap. J. {\bf 517}, 565.

\item[] J. Ponce de Leon, 1988, J. Math. Phys. {\bf 29}, 2479.

\item[] E. Poisson, 1990,  Phys. Rev. D  {\bf 41}, 1796.

\item[] P. J. Quinn et al., 1986,  Nature {\bf 322}, 329.\item[] L. Randall and R. Sundrum, 1999, Phys. Rev. Letts. {\bf
83}, 3370 \& 4690.

\item[] B. Ratra and P.J.E.Peebles, 1988,  Phys. Rev. D  {\bf 37}, 3406.
\item[] A.  D.  Rendall, 1997, Classical Quantum Gravity {\bf14}, 2341.
\item[] A.D. Rendall, 2003, Class. Quant. Grav. {\bf 21} 2445  [gr-qc/0312020].

\item[] A. G. Riess et al., 1998, Astron. J. {\bf 116}, 1109.
\item[] A. G. Riess et al., 2004, Astrophys. J.
{\bf 607}
 665 [astro-ph/0402512].

\item[] H. Ringstr\"{o}m,      2000, Classical Quantum Gravity {\bf 17}, 713.

\item[] H.~Ringstr\"{o}m, 2001,
Annales Henri Poincar\'e {\bf 2}, 405.

\item[] H. Ringstrom, 2003, Class. Quant. Grav. {\bf 20}, 1943.

\item[] M. D. Roberts, 1989, Gen. Rel. Grav. {\bf 21}, 907.\item[]  V.  Rubakov and M.  Shaposhnikov, 1985, Phys. Lett. B{\bf
159}, 22.

\item[] H. Sato, 1984, in {\it General Relativity and Gravitation}, ed. B. Bertotti, p. 289 (Reidel).
\item[] J. Schwartz, J. P. Ostriker and A. Yahil, 1975, Ap. J. {\bf 202},1.\item[] L. I. Sedov, 1967, {\it Similarity and Dimensional Methods inMechanics} (New York, Academic).
\item[] M. Shibata and M. Sasaki, 1999, Phys. Rev. D {\bf 60}, 084002.
%\item[] Y. Suto et al., 1995, Prog. Theor. Phys. {\bf 93}, 839.\item[] G. I. Taylor, 1950, Proc. Roy. Soc. London {\bf A201}, 175.\item[] R. C. Tolman, 1934, Proc. Nat. Acad. Sci. {\bf 20}, 169.
\item[] K. Tomita, 1995, Astrophys. J. {\bf 451}, 1.
\item[] K. Tomita, 1997, Phys. Rev. D{\bf 56}, 3341.
\item[] C.~Uggla, H.~van Elst, J.~Wainwright and G.F.R.~Ellis, 2003,
Phys.  Rev. D {\bf 68}, 103502.
\item[] H. van Elst and C. Uggla, 1997, Class. Quant. Grav. {\bf 14},  2673.

\item[]  R. J. van den Hoogen, A. A. Coley and Y. He, 2003,
 Phys. Rev. D. {\bf 68}, 023502.

\item[] J. Wainwright and L. Hsu, 1989, Class. Quantum Grav. {\bf 6},1409.\item[] J. Wainwright and G. F. R. Ellis (WE), 1997, {\it Dynamical systems incosmology} (Cambridge University Press, Cambridge).

\item[] J. Wainwright, A.A. Coley, G.F.R. Ellis and
M. Hancock,  1998,
Class.
Quant. Grav. {\bf 15} 331.

\item[] J.~Wainwright, M.J.~Hancock and C.~Uggla, 1999,
Class. Quant. Grav. {\bf 16}, 2577.

 \item[] R. M. Wald, 1983, Phys. Rev. D. {\bf 28}, 2118.

 \item[] A. Wang and Y. Wu, 2005, astro-ph/0504451.

\item[] B. Waugh and K. Lake, 1988, Phys. Rev. D {\bf 38}, 1315. \item[] B. Waugh and K. Lake, 1989, Phys. Rev. D {\bf 40}, 2137.\item[]  M. Weaver, 2000, Class. Quant. Grav. {\bf 17},  421.\item[] M.  Weaver, J.  Isenberg and B.  K.  Berger, 1998, Phys.  Rev.  Lett.{\bf 80}, 2984.

\item[] P. S. Wesson, 1989, Ap. J. {\bf 336}, 58.

\item[] C.  Wetterich, 1988, Nucl.  Phys.  B {\bf 302}, 66.  \item[] A.  Whitworth and D.  Summers,
1985, MNRAS, {\bf 214}, 1.

 %\item[] Wu, 1981, Gen.  Rel.  Grav.

 \item[] Ya.  B.  Zeldovich and A.
S.  Kompaneets, 1950, {\it Collection Dedicated to Joffe} {\bf 61}, ed.  P.I.  Lukirsky (Izd.  Akad.
Nauk SSSR, Moscow).

\item[] Ya.  B.  Zeldovich and Yu.  P.  Raizer, 1963, {\it Physics of Shock
Waves and High Temperature Phenomena} (New York, Academic).

\item[] Ya.  B.  Zeldovich and I.  D.
Novikov, 1967, Soviet Astr.  AJ.  {\bf10}, 602.

\end{enumerate}

\end{document}